\documentstyle[epsf,12pt]{article}
\begin{document}

\catcode`@=11
\long\def\@caption#1[#2]#3{\par\addcontentsline{\csname
  ext@#1\endcsname}{#1}{\protect\numberline{\csname
  the#1\endcsname}{\ignorespaces #2}}\begingroup
    \small
    \@parboxrestore
    \@makecaption{\csname fnum@#1\endcsname}{\ignorespaces #3}\par
  \endgroup}
\catcode`@=12
\newcommand{\newc}{\newcommand}
\newc{\lat}{{\ell at}}
\newc{\mgut}{M_{\rm GUT}}
\newc{\five}{{\bf 5}}
\newc{\fivebar}{{\bf\bar 5}}
\newc{\ten}{{\bf 10}}
\newc{\tenbar}{{\bf\bar{10}}}
\newc{\sixteen}{{\bf 16}}
\newc{\sixteenbar}{{\bf\bar{16}}}
\newc{\gsim}{\lower.7ex\hbox{$\;\stackrel{\textstyle>}{\sim}\;$}}
\newc{\lsim}{\lower.7ex\hbox{$\;\stackrel{\textstyle<}{\sim}\;$}}
\newc{\gev}{\,{\rm GeV}}
\newc{\mev}{\,{\rm MeV}}
\newc{\ev}{\,{\rm eV}}
\newc{\kev}{\,{\rm keV}}
\newc{\tev}{\,{\rm TeV}}
\newc{\mz}{m_Z}
\newc{\mw}{m_W}
\newc{\mpl}{M_{Pl}}
\newc{\mh}{m_h}
\newc{\mA}{m_A}
\newc{\chifc}{\chi_{{}_{\!F\!C}}}
\newc\order{{\cal O}}
\newc\CO{\order}
\newc\CL{{\cal L}}
\newc\CY{{\cal Y}}
\newc\CH{{\cal H}}
\newc\CM{{\cal M}}
\newc\CF{{\cal F}}
\newc\CD{{\cal D}}
\newc\CN{{\cal N}}
\newc{\eps}{\epsilon}
\newc{\re}{\mbox{Re}\,}
\newc{\im}{\mbox{Im}\,}
\newc{\invpb}{\,\mbox{pb}^{-1}}
\newc{\invfb}{\,\mbox{fb}^{-1}}
\newc{\yddiag}{{\bf D}}
\newc{\yddiagd}{{\bf D^\dagger}}
\newc{\yudiag}{{\bf U}}
\newc{\yudiagd}{{\bf U^\dagger}}
\newc{\yd}{{\bf Y_D}}
\newc{\ydd}{{\bf Y_D^\dagger}}
\newc{\yu}{{\bf Y_U}}
\newc{\yud}{{\bf Y_U^\dagger}}
\newc{\ckm}{{\bf V}}
\newc{\ckmd}{{\bf V^\dagger}}
\newc{\ckmz}{{\bf V^0}}
\newc{\ckmzd}{{\bf V^{0\dagger}}}
\newc{\X}{{\bf X}}
\newc{\bbbar}{B^0-\bar B^0}
\def\bra#1{\left\langle #1 \right|}
\def\ket#1{\left| #1 \right\rangle}
\newc{\sgn}{\mbox{sgn}\,}
\newc{\m}{{\bf m}}
\newc{\msusy}{M_{\rm SUSY}}
\newc{\munif}{M_{\rm unif}}
\newc{\slepton}{{\tilde\ell}}
\newc{\Slepton}{{\tilde L}}
\newc{\sneutrino}{{\tilde\nu}}
\newc{\selectron}{{\tilde e}}
\newc{\stau}{{\tilde\tau}}
%
%
\def\NPB#1#2#3{Nucl. Phys. {\bf B#1} (19#2) #3}
\def\PLB#1#2#3{Phys. Lett. {\bf B#1} (19#2) #3}
\def\PLBold#1#2#3{Phys. Lett. {\bf#1B} (19#2) #3}
\def\PRD#1#2#3{Phys. Rev. {\bf D#1} (19#2) #3}
\def\PRL#1#2#3{Phys. Rev. Lett. {\bf#1} (19#2) #3}
\def\PRT#1#2#3{Phys. Rep. {\bf#1} (19#2) #3}
\def\ARAA#1#2#3{Ann. Rev. Astron. Astrophys. {\bf#1} (19#2) #3}
\def\ARNP#1#2#3{Ann. Rev. Nucl. Part. Sci. {\bf#1} (19#2) #3}
\def\MPL#1#2#3{Mod. Phys. Lett. {\bf #1} (19#2) #3}
\def\ZPC#1#2#3{Zeit. f\"ur Physik {\bf C#1} (19#2) #3}
\def\APJ#1#2#3{Ap. J. {\bf #1} (19#2) #3}
\def\AP#1#2#3{{Ann. Phys. } {\bf #1} (19#2) #3}
\def\RMP#1#2#3{{Rev. Mod. Phys. } {\bf #1} (19#2) #3}
\def\CMP#1#2#3{{Comm. Math. Phys. } {\bf #1} (19#2) #3}
\relax
%
%
%
\def\beq{\begin{equation}}
\def\eeq{\end{equation}}
\def\bea{\begin{eqnarray}}
\def\eea{\end{eqnarray}}
%
%
%
\newc{\ie}{{\it i.e.}}          \newc{\etal}{{\it et al.}}
\newc{\eg}{{\it e.g.}}          \newc{\etc}{{\it etc.}}
\newc{\cf}{{\it c.f.}}
\def\smuon{{\tilde\mu}}
\def\neut{{\tilde N}}
\def\char{{\tilde C}}
\def\bino{{\tilde B}}
\def\wino{{\tilde W}}
\def\higgsino{{\tilde H}}
\def\sneut{{\tilde\nu}}
%
%
%
%
\def\slash#1{\rlap{$#1$}/} 
\def\Dsl{\,\raise.15ex\hbox{/}\mkern-13.5mu D} 
\def\delsl{\raise.15ex\hbox{/}\kern-.57em\partial}
\def\Ksl{\hbox{/\kern-.6000em\rm K}}
\def\Asl{\hbox{/\kern-.6500em \rm A}}
\def\Qsl{\hbox{/\kern-.6000em\rm Q}}
\def\gradsl{\hbox{/\kern-.6500em$\nabla$}}
%
%
%
\def\bar#1{\overline{#1}}
\def\vev#1{\left\langle #1 \right\rangle}
%

\begin{titlepage} 
\begin{flushright}
OSU-HEP-04-11\\
October 2004\\
\end{flushright}
\vskip 2cm
\begin{center}
{\large\bf 
Perturbative Unification and Higgs Boson Mass Bounds
}
\vskip 1cm
{\normalsize\bf
K.S.~Babu$^{(a)}$, Ilia Gogoladze$^{(b)}$ and Christopher Kolda$^{(b)}$ \\
\vskip 0.5cm
$^{(a)}${\it Oklahoma Center for High Energy Physics, 
Department of Physics, \\ Oklahoma State University,
Stillwater, OK~~74078, USA}\\ 
~~ \\
$^{(b)}${\it Department of Physics, University of Notre Dame\\
Notre Dame, IN~~46556, USA}\\[0.1truecm]
}

\end{center}
\vskip .5cm

\begin{abstract}
Supersymmetric extensions of the Standard Model generally give a
theoretical upper limit on the 
lightest Higgs boson mass which may be uncomfortably close to the
current experimental lower bound
of $m_h > 114$ GeV.  Here we show ways in which this upper limit on
$m_h$ can be relaxed substantially 
in the perturbative regime, while maintaining the successful
unification of gauge couplings as observed in the 
minimal scenario.  A class of models is presented, which includes new
vector-like matter with the quantum numbers 
of the ${\bf 5}$, ${\bf\bar 5}$ and singlet representations of $SU(5)$
and having Yukawa couplings ($\kappa$) 
to the usual Higgs doublet $H_u$.  This new matter transforms
non-trivially under a ``lateral" $SU(N)$ 
gauge symmetry, which enables the new Yukawa couplings $\kappa$ to be
sizable, even larger than the 
top-quark Yukawa coupling, and still be perturbative at all momenta
up to the unification scale of $10^{16}\,$GeV.  
We find that, consistent with low energy constraints, $m_h$ can
be raised to values as large as 200 -- $300\,$GeV. 
 
\end{abstract}

\end{titlepage}

\setcounter{footnote}{0}
\setcounter{page}{1}
\setcounter{section}{0}
\setcounter{subsection}{0}
\setcounter{subsubsection}{0}


With the discovery nearly a decade ago of the top quark, the
penultimate piece of the Standard Model (SM), our full attention has finally
turned to the missing link -- the Higgs boson. While its discovery
would complete the SM, it would also bring to the forefront
the study of electroweak
symmetry breaking and all that comes with it
(supersymmetry, extra dimensions, etc). But discovery of the Higgs
boson is no simple feat. In particular, the channels in which it can
be observed, and thus the methods used to find it, are highly
sensitive to its mass, which remains unknown. Though there do exist
constraints on $m_h$ from loop processes in the SM, 
the dependence on $m_h$ is logarithmic and thus small uncertainties in
electroweak observables lead to exponentially larger uncertainties in
the Higgs mass itself.

Within the SM, the mass of the Higgs boson is not predicted. While present
analyses of the precision radiative constraints within the SM indicate
a Higgs mass somewhere around 100~GeV, the model itself puts only weak
constraints on $m_h$ from unitarity: $m_h\lsim
700\gev$~\cite{unitarity}. Even after we embed the model in
a more complete ultraviolet picture, the added constraints of vacuum
stability and triviality only serve to loosely bound the Higgs mass
range~\cite{triviality,kolda}.

The minimal supersymmetric extension of the SM (the MSSM) is another
story. It is well known that the mass of the lightest Higgs boson of
the MSSM is bounded at tree level by the $Z^0$ mass. Radiative
corrections can push the mass higher, but only to about 125 to $130\gev$ for
$\msusy\lsim 1\tev$. (Here $\msusy$ represents the average mass of the top
squarks.) As experiments have pushed the lower mass bound on the Higgs
higher and higher, the parameter space for SUSY has been pushed
further and further into corners in which top squarks are heavy and
left-right mixing is large. In essence, even small increases in the
experimental bound on $m_h$ lead to large increases in the
tuning required to make the MSSM consistent with experiment~\cite{little}.

Many groups have examined ways to extend the MSSM in order to push up
the lightest Higgs mass and partially evade the tuning
issues~\cite{littleideas}. 
Generally speaking, the farther one is willing to move away from
the basic picture of the MSSM, the higher one is able to push the
Higgs mass. Models which add only a few extra matter
representations, or additional Higgs multiplets, succeed only in
raising the Higgs mass bound by a few to few tens of GeV. However
there are models which look nothing like the MSSM in the ultraviolet,
involving for example new strong dynamics in which the MSSM degrees of
freedom are no longer fundamental. In such models the upper bound on
$m_h$ can be extended by hundreds of GeV.

In this paper, we would like to keep the best of both worlds --
minimality but with a large effect. We will
present a model (actually a class of models) in which very large
corrections to the Higgs mass can occur, while remaining as close as
possible to the spectrum and interactions of the MSSM. In particular,
we wish to preserve a property of the MSSM which we consider too
valuable to abandon, namely gauge coupling unification which occurs at
scales sufficiently large to prevent fast proton decay. 

This last property is really two independent requirements, both of
which are met in the MSSM: first, that the gauge couplings unify
numerically and perturbatively, and second that they unify at a
sufficiently large scale that the resulting dimension-6 nucleon decay
rates are consistent with proton lifetime bounds. If we are to
consider the unification in the MSSM as more than a coincidence then
there are only a few possible extensions which we may consider: the
addition of gauge singlets fields, as in the NMSSM~\cite{nmssm,morehiggs}; 
the addition of new matter fields whose quantum numbers fit into 
complete SU(5) representations~\cite{su5}; or the addition of
new interactions commuting with the usual gauge groups~\cite{erler}.

Of these, the first two have been considered in detail by others and
the last only in passing (since it usually does not help). 
We consider all three in tandem and show that we can generate Higgs
masses far above those predicted in the MSSM. Because the quantum
numbers of the new matter will allow it to couple to the $H_u$ field,
the new Yukawa couplings will generate large corrections to the light
Higgs mass in much the same way that the top quark/squark loops do in
the MSSM. But as we will see, the Yukawa coupling can be appreciably
larger than the top Yukawa of the MSSM, consistent with perturbative
unification, and so the corrections can be extremely large.

\section*{Higgs Mass in the MSSM}

Before discussing our model, 
let us briefly review the calculation of $\mh$ in the MSSM and its
concomitant problems. At tree level, the lightest scalar Higgs mass is
bounded from above by $\mh\leq \mz\cos2\beta$. The bound is saturated
in the decoupling limit where the pseudoscalar Higgs mass is taken to infinity:
$m_A\to\infty$. In determining the upper bound in $m_h$, 
we will take this limit for the remainder of the paper. 

There are several large corrections to this bound, but they are
essentially dominated by terms in the effective potential of the form
$(H_u^\dagger H_u)^2$. These corrections have been calculated by a
large number of authors and reviewed already in several other
papers~\cite{at, Carena:1995wu}. 
Nonetheless we remind the reader of the leading 1- and 2-loop
contributions:
\begin{eqnarray}
\mh^2& = & \mz^2\cos^2
2\beta\left( 1-\frac{3}{8\pi^2}\frac{m_t^2}
{v^2}\ t\right) \label{mssm} \\
 & + & \frac{3}{4\pi^2}\frac{m_t^4}{v^2}\left[
t+\frac{1}{2}X_t +
\frac{1}{16\pi^2}\left(\frac{3}{2}\frac{m_t^2}{v^2}
-32\pi\alpha_s \right)\left(X_t
t+t^2\right)\right]. \nonumber
\end{eqnarray}
Here $v=174\gev$ is the usual SM Higgs vev and $t=\log(\msusy^2/m_t^2)$.
The first term in square brackets is the famously large ($\sim m_t^4$)
correction to the light Higgs mass, while the first term in
parenthesis is a smaller D-term contribution.
The terms proportional to $X_t$ are corrections due to 
left-right top squark mixing. Here $X_t$ is defined as
\beq
X_t =\frac{2 \tilde A_t^2}{\msusy^2}\left (1-\frac{\tilde
A_t^2}{12 \msusy^2}\right)
\eeq
where $\tilde A_t=A_t-\mu\cot\beta$ is the top squark mixing parameter.
The remaining terms in Eq.~(\ref{mssm}) are the leading 2-loop
effects, which we will keep throughout this analysis.

The size of the corrections is logarithmically dependent on the
``cut-off'' which is here the mass scale of the top squarks:
$\msusy\simeq m^2_{\tilde t_1,\tilde t_2}$. For simplicity (and
following standard conventions) 
we will take all squarks to be 
degenerate, apart from mass splittings induced by mixing.
To find an upper bound on $\mh$ is it customary to take $\msusy$ as
large as possible consistent with our prejudices about fine
tuning. Usually this means taking $\msusy=1\tev$.
Using that value, one finds
for the unmixed case ($X_t=0$) that 
$\mh<110\gev$. The light Higgs mass is maximized for the case
$X_t=6$ (or $A_t=\sqrt{6}\msusy$)
where one find $\mh<125\gev$. These two extrema are usually referred to
as the minimal and maximal mixing scenarios. 

Current experimental constraints on $\mh$ already rule out a
substantial portion of the mass range of the light
Higgs. For large $m_A$, the SM Higgs bound translates
directly over to the MSSM and one has $\mh>114\gev$ at 90\%
CL~\cite{pdg}. 

One can turn this around and ask how large $\msusy$ must be in order
to conform to experiment. For example, if $A_t=0$ 
one finds that $\msusy\geq 1.5\tev$. Such a large value for $\msusy$
will induce rather large corrections (of $O(\tev)$) for the mass
parameters in the Higgs potential. In order to generate a vev of
$174\gev$, one must assume a delicate cancellation, either between the
$\mu$-term and the Higgs mass parameters, or between the bare mass
parameters and their one-loop corrections~\cite{little}. 
In either guise, this
tuning (usually called the ``little hierarchy problem'') 
has led many theorists to consider ways to extend
the MSSM in order to push up $m_h$~\cite{littleideas}.

We will be doing something similar but not exactly the same. Our goal
is not to solve this little hierarchy problem, but rather to ask how
far we can push the Higgs mass consistent with phenomenological bounds
and gauge coupling unification. Some of the parameter space we are
considering will require fine-tuning as in the little hierarchy
problem, some will not, as we will discuss later.

\section*{The Model: Extra Matter}

It has long been understood that one can extend the matter sector of
the MSSM and still preserve the beautiful result of gauge coupling
unification if the new matter falls into complete multiplets
of SU(5). That is, the quantum numbers of the new matter must exactly
fill up representations of SU(5); there is no need for SU(5) to be
respected by the interactions, just by the matter content. 
Such complete, but light, GUT multiplets are not unexpected. Within
string theory one often finds light multiplets in the spectrum. And
even within the framework of GUTs themselves one can find extra complete
multiplets lying at the weak scale~\cite{Babu:1996zv}. In particular,
if a symmetry prevents a multiplet from receiving a GUT-scale mass,
that multiplet will then usually get its mass from SUSY-breaking
effects, which naturally places it at the weak scale.

However there are stringent constraints on new light matter from a
number of sides. Most important are the constraints from the $S$ and
$T$ parameters which limit the number of additional {\it chiral}\/
generations.
Consistent with these constraints, one must add new matter which is
predominantly vector-like and only has a chiral mass which is subdominant.

Since the smallest representations of SU(5) are chiral, 
we must therefore introduce them in pairs (\eg, $\five+\fivebar$ or $\ten
+\tenbar$) in order to allow 
explicit mass terms. But we must also allow for the possibility that
these fields couple to the Higgs fields of the MSSM.
For example, with a $\ten+\tenbar$ a mass term of
the form $\ten\cdot\ten\cdot H_u$ is allowed\footnote{By this notation we
actually mean a mass term for the up-quark-like member of the $\ten$,
not the entire $\ten$, since $H_u$ represents only a part of a
complete $\five$ representation. We use a similar shortcut when we
write $\fivebar\cdot S\cdot H_u$ and $\five\cdot S\cdot H_d$.}. 
If we allow SM singlets, then Dirac mass terms
of the form $\fivebar\cdot S\cdot H_u$ and $\five\cdot S\cdot H_d$ 
are also permitted.

How large can the new Yukawa couplings be? There are two severe constraints.
The first comes from the $T$ parameter. In the limit $M_V\gg M_D$, the
contribution from a single chiral fermion is  
approximately~\cite{Lavoura:1992np}:
\begin{equation}
T=\frac{ N (\kappa v)^2}{10 \pi \sin^2\theta_W m^2_W}\left[ \left( 
\frac{\kappa
v}{M_V}\right)^2  +O\left( \frac{\kappa v}{M_V}\right)^4\right] 
\label{tpar}
\end{equation}
where $\kappa$ is the new Yukawa coupling, $v$ is vev of corresponding
Higgs field, and $N$ counts the additional degrees of freedom possessed
by the new field. Precision electroweak data constrains $T=-0.17\pm
0.12$ for $m_h=117\gev$ or $-0.08\pm0.12$ for
$m_h=300\gev$~\cite{pdg}. As a figure of merit, we will take $T<0.1$
as a realistic bound and apply it in our analysis.
Then it is easy to see from Eq.~(\ref{tpar}) that when
$M_V$ is around 1~TeV the Yukawa coupling $\kappa$ can be $O(1)$!


The second constraint comes from perturbativity and unification. 
If we are to
maintain natural gauge coupling unification, then we must also
maintain perturbativity of the gauge and Yukawa couplings at all
scales below $\mgut$. A new $O(1)$ coupling presents a problem both in
its own evolution towards the GUT scale, but also in its effect on the
running of the top quark Yukawa, which is already known to be near its
quasi-fixed point. Since the self-coupling of any new Yukawa would
drive that same Yukawa up in the ultraviolet, and may also drive up
$y_t$, the new Yukawa coupling cannot be too large.
For example, the $\ten\cdot\ten\cdot H_u$ coupling cannot be larger
than $1.1$. Even worse, the $\fivebar\cdot S\cdot H_u$ coupling cannot be
larger than $0.75$.

Perturbativity and unification also bound the amount of new matter
than can be added to the MSSM, since each new
representation {\it increases}\/ the gauge $\beta$-functions.
One finds that one can safely add to the low-energy (TeV-scale)
spectrum only the following: 
{\it (i)} up to 4 pairs of $(\five+\fivebar)$'s, 
or {\it (ii)} one pair of $(\ten+\tenbar)$, or {\it (iii)} one pair
of each,
$(\five+\fivebar+\ten+\tenbar)$. The last option of course also fits
nicely into an SO(10) model. In addition, any number of gauge singlets
can be added without upsetting unification or perturbativity.

Cases (ii) and (iii) have been studied extensively before. For example,
Moroi and Okada~\cite{su5} concluded 
that the mass of the lightest CP even Higgs mass could be pushed up as
high as 160~GeV consistent with all perturbativity constraints.

By itself, case (i) does not allow for any new Yukawa coupling unless
the new states in the $\fivebar$ are mixed into the usual $d$-quarks
and leptons. Such a possibility is even more strongly constrained (by
flavor violation and 
unitarity of the CKM matrix, among others) and so we will forbid all
such mixing. In the next section an explicit argument for doing so
will appear.

But we can generate new Yukawa couplings for case (i) if we introduce
gauge singlet fields $S$ and $\bar S$ which couple to the usual Higgs
bosons as $\fivebar\cdot \bar S\cdot H_u$ and $\five\cdot S\cdot H_d$.
More explicitly, we decompose the $\five$ as $(d',\bar L')$ and
$\fivebar$ as $(\bar d',L')$,
where $d'$ and
$L'$ have the same quantum numbers as the $d_R$-quarks and $L_L$
leptons of the SM. Then we generate a superpotential with only two new
Yukawa interactions. For one pair of $\five+\fivebar$ and one
pair of singlets $S+\bar S$, the superpotential is:
\begin{equation}
W=\kappa L' \bar S H_u + \kappa' \bar L' S H_d+ 
M_V(\bar SS + \bar L'L'+\bar d'd')
\label{yuk5}
\end{equation}
where we have taken a common vector-like mass for simplicity.
Thus the lepton-like pieces of the
$\five$ and $\fivebar$ get Dirac {\it and}\/ vector-like 
masses, while leaving the
$d$-quark-like pieces with only vector-like masses.

Because the new matter couples to the SM Higgs fields, they will
generate radiative corrections in much the same way as the top
quark and squarks. And like the top/bottom sector, the coupling to
$H_u$ will play a much more dominant role in those corrections than
the coupling to $H_d$, assuming $\tan\beta>1$. This is important
because both the $\kappa$ and $\kappa'$ couplings will tend to run to
large values in the ultraviolet. For example, if we have $N$ pairs of
$\five +\fivebar$ and if all couple with the same strength $\kappa$ 
then:\footnote{Throughout this work, we use full 2-loop RGEs to run
  the parameters of the model, including $\kappa$ and $y_t$. However
  we only show the 1-loop contributions here in order to be concise.}
\begin{equation}
\frac{d}{d\log Q} \kappa =\frac{\kappa}{16\pi} \left[(3+N) \kappa^2
  +3 y_t^2
  -\left( \frac{3}{5} g_1^2+3 g_2^2 \right) \right].
\label{rge11}
\end{equation}
Of perhaps even greater importance is the RGE for $y_t$, which has a
new contribution proportional to $\kappa^2$:
\beq
\frac{d}{d\log Q}y_t = \left[\frac{d}{d\log Q}y_t\right]_{\rm MSSM} \!
+\frac{N}{16\pi^2} y_t\kappa^2.
\eeq
Because $\kappa$ and $y_t$ are coupled through their RGEs, 
$\kappa$ is more tightly constrained than $\kappa'$. 
However the mass of the
lightest Higgs, $h^0$, is much more sensitive to $\kappa$ than it is to
$\kappa'$. Furthermore, very large values of $\kappa'$ will actually
drive the Higgs mass {\it down}. Therefore we will take $\kappa'=0$ for
the remainder of this work. 

Running the RGEs for $\kappa$ and $y_t$ from the weak scale to the GUT
scale (which we take as $2\times10^{16}\gev$), we find an upper bound
on $\kappa$ of $0.75$ for $N=1$ and $0.65$ for $N=3$ (assuming
$\tan\beta\gg1$). 
Calculating the correction to the Higgs mass (formulas to follow in
next section) yields a new upper bound on $m_h$ which is at most $20\gev$
above the MSSM limit, not much of an improvement.

Of course the correction to $\mh^2$ scales as $\kappa^4$, so if we
were able to significantly increase the low-energy value of $\kappa$
we could generate much larger Higgs masses, as we will now see.

\section*{The Lateral Gauge Symmetry}

Models with extra pairs of $(\five+\fivebar)$'s present one extra degree
of freedom in model building. Specifically, if one has $N$ pairs of
$\five+\fivebar$ then one can impose on them a new $SU(N)$
symmetry under which the SM particles are uncharged. 
As a global symmetry, this $SU(N)$ only serves to constrain
the forms of the couplings; for example, it would prevent mixing
between the vector-like quarks and the ordinary quarks unless the
symmetry is broken.

As a gauge symmetry, there are new
opportunities for generating large corrections to $m_h$. Because it is
reminiscent of ``horizontal'' symmetries among the quark generations,
we will call this a ``lateral'' symmetry and denote it $SU(N)_\lat$.
Because of perturbativity there are only three possible lateral
symmetry groups: $SU(2)$, $SU(3)$ and $SU(4)$. ($Sp(4)$ is another
possible group, but we will not consider it.) We will keep most of
our discussion general with respect to $N$, but our numerical work
will be done for $N=3$ just to be specific.

The non-MSSM matter content is then as follows, with the charges under
$SU(3)_c\times SU(2)_W\times SU(N)_\lat \times U(1)_Y$ shown:
\begin{eqnarray}
d'\,({\bf 3},{\bf 1},{\bf N})_{1/3},&&
\bar d'\,({\bf\bar 3},{\bf 1},{\bf \bar N})_{-1/3},\nonumber \\ 
L'\,({\bf 1},{\bf 2},{\bf N})_{1/2},&&
\bar L'\,({\bf 1},{\bf 2},{\bf\bar N})_{-1/2},\label{matter}\\
S\,({\bf1},{\bf1},{\bf N})_0,&&
\bar S\,({\bf 1},{\bf1},{\bf\bar N})_0. \nonumber
\end{eqnarray}
The superpotential is again given by Eq.~(\ref{yuk5}). Again we
will take $\kappa'\to 0$, which has very little effect on our limits.

The major advantage of gauging the lateral symmetry comes in the
RGEs for $\kappa$. Let us reiterate that in order to maximize the
Higgs mass correction, we want $\kappa$ to be as large as possible at
the weak scale. However a large $\kappa$ enters the RGE for $y_t$,
driving it to an ultraviolet instability long before the GUT scale.
Furthermore the $\beta$-function for 
$\kappa$ itself is positive, which means that a large $\kappa$ in the
infrared only becomes larger, and more troublesome, in the ultraviolet.

Gauging the lateral symmetry changes all this because now one finds:
\begin{equation}
\frac{d}{d\log Q} \kappa =\frac{\kappa}{16\pi} \left[(3+N) \kappa^2 
+3 y_t^2 -\left( \frac{3}{5} g_1^2+3 g_2^2 +4 C_2^\lat g_{\lat}^2 
\right) \right] 
\label{rge1}
\end{equation}
where $C_2^\lat$ is the quadratic Casimir for the fundamental representation
of $SU(N)_\lat$:
$C_2^\lat=(N^2-1)/2N$. 
The RGE for the new coupling $g_\lat$ or $\alpha_\lat$ is simply:
\beq
\frac{d}{d\log Q}\alpha_\lat = \frac{1}{2\pi}\beta_\lat\,\alpha^2_\lat
\eeq
where $\beta_\lat$ is the numerical $\beta$-function coefficient, to
be determined.

\begin{figure}
\centering
\epsfysize=3in
\hspace*{0in}
\epsffile{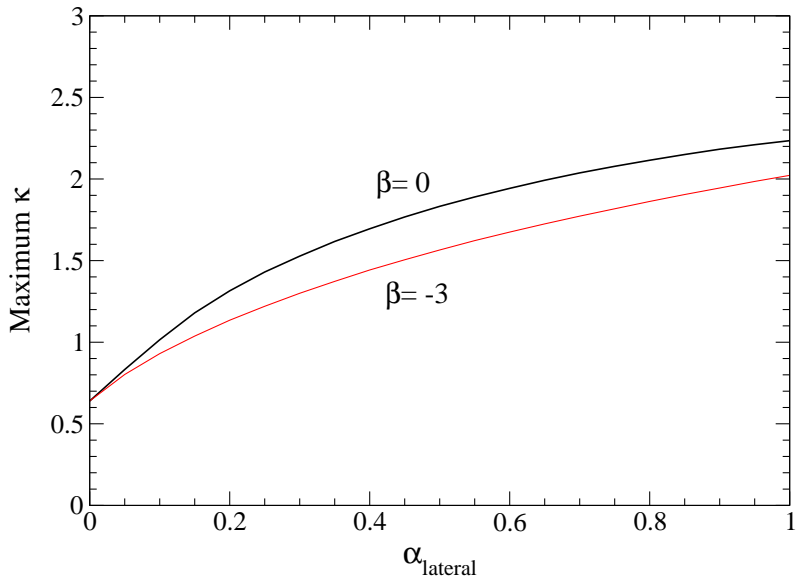}
\caption{Plot of maximum allowed $\kappa$ as a function of
  $\alpha_\lat$ for the $N=3$ case with $\beta_\lat = 0$ and $-3$.}
\label{kappa}
\end{figure}

How does $\alpha_\lat$ help? In Eq.~(\ref{rge1}) it is the gauge
pieces that drive down $\kappa$ in the ultraviolet, 
preventing $\kappa$ from hitting its Landau pole. More importantly they
suppress the contributions of $\kappa$ to the running of $y_t$. Thus
much larger values of $\kappa$ are possible for non-zero
$\alpha_\lat$. In Fig.~\ref{kappa} we graph the upper bound on
$\kappa$ as a function of $\alpha_\lat$ for $N=3$ and two cases of
$\beta$: $\beta_\lat=0$ and $-3$. 
Just as the top quark corrections to $m_h^2$ scale as $y_t^4$, so too
these new corrections scale as $\kappa^4$. So it would appear that
larger $\alpha_\lat$ immediately leads to larger $m_h$, and to a good
approximation this is correct, though not complete.

We are finally in a position to give the expression for the radiative
correction to $m_h^2$ coming from the new matter:
\begin{eqnarray}
\mh^2& = & \left[ m_h^2\right]_{MSSM}- \mz^2\cos^2
2\beta\left(\frac{N}{8\pi^2}\kappa^2\ t_1\right) \label{lat2} 
\label{hm2}\\
& + & N \kappa^4 \frac{v^2 \sin^4 \beta}{4\pi^2}\left[
t_1+\frac{1}{2}X_\kappa +
\frac{1}{16\pi^2}\left(\frac{N}{2}\kappa^2
-6 C_2^\lat g^2_\lat \right)\left(X_\kappa
t_1+t_{1}^2\right)\right]. \nonumber
\end{eqnarray}
where 
\beq
t_1=\log\left(\frac{M^2_{SUSY}+M_V^2}{M_V^2}\right)
\eeq
and we have assumed $M_V\gg M_D$. Here $X_\kappa$ is
analogous to $X_t$ in the stop sector:
\beq
X_\kappa 
=\frac{2 \tilde A_\kappa^2}{\msusy^2+M_V^2}\left (1-\frac{1}{12}\frac{\tilde
A_{\kappa}^2}{\msusy^2+M_V^2}\right)
\label{X}
\eeq
where $\tilde A_{\kappa}=A_{\kappa}-\mu\cot\beta$ is the $L'$--$\bar S$
soft mixing parameter.

We learn several things from the form of Eq.~(\ref{hm2}). First,
it is clear that the corrections grow as $\kappa^4$, so large $\kappa$
will generate large Higgs masses. Second, we notice that the 2-loop
piece contains terms suppressed by $\alpha_\lat$, so for sufficiently large
$\alpha_\lat$, the 2-loop suppression of $m_h^2$ offsets the
enhancements generated by the $\alpha_\lat$-dependence in the RGEs.

We have outlined here a broad class of models which can
significantly increase the SUSY light Higgs mass. Each model is
described by only a few free parameters: $N$, the number of
$(\five+\fivebar)$ multiplets; $M_V$, the explicit, SUSY-preserving
mass of those multiplets; $\msusy$, the soft-breaking mass scale;
$\kappa$, the Yukawa coupling of the new matter to $H_u$;
$\alpha_\lat$, the strength of the new $SU(N)$ gauge interaction; 
$\beta_\lat$, the $\beta$-function for $\alpha_\lat$; and
of course $\tan\beta$, the ever-present ratio of the Higgs vevs.
As in the MSSM, there is also a dependence on the left-right mixing
parameters, $\tilde A_t$ and $\tilde A_\kappa$, which explicitly 
appear in the Higgs mass correction.

In what follows we will study one particular subclass of models in
order to show the power of this approach for lifting the Higgs mass
predictions.

\section*{The $SU(3)_\lat$ Case}

Now we turn our attention to one choice of $N$, namely $N=3$. Our
choice is not dictated by any exceptionally deep reason. One could
argue that the $N=3$ case fits nicely into an $E_6$ unified picture
(where there is an extra pair of $\five+\fivebar$ in each {\bf
  27}) and is therefore preferred. But the $SU(3)_\lat$ case will also
highlight all the key points we wish to make.

The matter content (Eq.~(\ref{matter})) and superpotential (Eq.~(\ref{yuk5}))
of the $SU(3)_\lat$ model both
follow from the discussion in the previous sections. As before we will
make two simplifying assumptions: that all new matter receives a
universal vector-like mass $M_V$ and that all SUSY scalars receive a
universal scalar mass $\msusy$. Deviations from these assumptions will
have only a negligible effect on the derived bound on $m_h$.

What is not immediately obvious is what we should take for
$\beta_\lat$. The minimal model yields $\beta_\lat=-3$. In such a
model all the matter charged under $SU(3)_\lat$ would be integrated
out at the scale $M_V$. In the infrared the only remaining degrees of
freedom would be gauge and would thus form ``lateral glueballs'' which
would not have direct interactions with SM matter.

On the other hand, it would be natural for the model to contains Higgs
fields to break the lateral symmetry. A model with three
Higgs fields, each in the fundamental of $SU(3)_\lat$, along with 
their conjugates would break the lateral symmetry completely. The
resulting model would have $\beta_\lat=0$. Of course, any additional 
matter would
turn $\beta_\lat>0$. 

This brings us to an interesting question: could
an Abelian symmetry work for us? The answer is no. 
As we will see shortly, at the weak scale we require the lateral
interaction to be fairly strong in order to stabilize the RGE flow of
$\kappa$ and $y_t$. An Abelian symmetry
would have positive $\beta$-function and so the strongly coupled
lateral group would become non-perturbative in the ultraviolet.

\begin{figure}
\centering
\epsfysize=2.75in
\hspace*{0in}
\epsffile{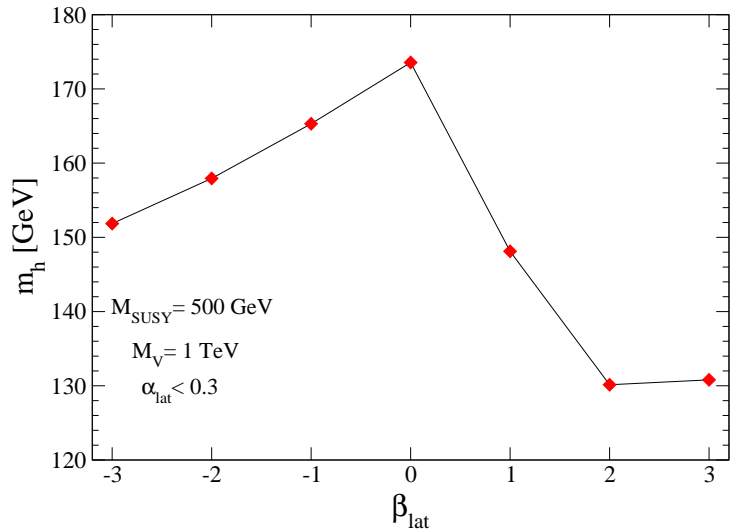}
\caption{Plot of maximum allowed Higgs mass, $m_h$, as a
  function of $\beta_\lat$ for $SU(3)_\lat$ with $\msusy=500\gev$,
  $M_V=1\tev$ and $\alpha_\lat(m_Z)<0.3$.} 
\label{betah}
\end{figure}

On the other hand, a negative $\beta$-function allows $\alpha_\lat$ to
be large in the infrared, but suppresses it in the ultraviolet so that
its dampening effect on the running of $\kappa$ is vitiated. Logically
it seems that $\beta_\lat=0$ maximizes the effect that $\alpha_\lat$
has on the Higgs mass calculation, a deduction borne out by the numerical
calculation shown in Fig.~\ref{betah} where
explicit dependence on the $\beta$-function is shown.
In this figure we have plotted the upper bound on $m_h$ as a
function of $\beta_\lat$ for $\msusy=500\gev$
and $M_V=1\tev$. We require in this one plot that
$\alpha_\lat(m_Z)<0.3$ and that $\alpha_\lat$  
remains perturbative all the way to
the GUT scale. The plot clearly shows $\beta=0$ as a maximum. However
all $\beta\leq0$ are interesting and give sizable enhancements to
$m_h$. Positive $\beta$-functions help very little,
since they drive $\alpha_\lat$ non-perturbative in the ultraviolet and
thus necessitate lower values of $\alpha_\lat$ at the weak scale.

Our central result is then the maximization of  
$m_h$ as a function of $\alpha_\lat$. We leave $\alpha_\lat$ as a free
variable since there is a strong dependence on its value. We also
leave $\msusy$ free since we have no {\it a priori}\/ reason to choose
$200\gev$ rather than $1\tev$. And we choose $\beta_\lat=0$ and $-3$.

How large
can $\alpha_\lat$ be before our perturbative calculation of the Higgs
mass begins to fail? Because we assume $\beta_\lat\leq 0$, there is no
problem with perturbativity in the ultraviolet. Furthermore, in the
calculation of the new contributions to $m_h^2$, the proper expansion
parameter is not $\alpha_\lat$, but rather $\alpha_\lat/4\pi$. This is
analogous to the QCD contributions to leptonic magnetic moments. Thus
our calculation of $m_h$ should work reasonably well even for values
of $\alpha_\lat$ much larger than $\alpha_{\rm QCD}(m_Z)$.

In any case, our numerical procedure is simple: For a given value of
$\alpha_\lat$ and $\msusy$, we vary $M_V$ and $\kappa$ over their
allowed ranges in order to maximize $m_h$. The minimum for $M_V$
is set by requiring $T<0.1$; this will of course depend on
$\kappa$. The upper bound for $M_V$ we choose to be $2\tev$
arbitrarily. While there is no lower bound on $\kappa$, the upper
bound is set by requiring the values of $y_t(Q)$ and $\kappa(Q)$
remain perturbative at all scales $Q\leq\mgut$.

We have thus far avoided discussing the left-right mixing
parameters. It turns out that these play a crucial role. It is well
known that in the MSSM, the Higgs mass is maximized for $X_t=6$, or
equivalently $\tilde A_t=\sqrt{6}\msusy$. For fields with both chiral
and vector-like masses, the form of $X$ is altered to that of Eq.~(\ref{X}).
The one-loop corrections are maximized for $X_\kappa=6$, however it
may not be possible to reach such large values.
In fact, there are really two distinct cases to be considered,
both of which are physically acceptable, though one may (or
may not) be considered more natural than the other.

The first case is to require that $A_\kappa$ scales with
$\msusy$. For example, in the MSSM 
the requirement that there be no charge- or color-breaking
minima in the scalar potential lower than the SM minimum gives us the
usual~\cite{ccb}:
\beq
A_t^2\leq 3\left(m^2_{\tilde t_L}+m^2_{\tilde t_R}+m^2_{H_u}\right).
\eeq
Since $m^2_{H_u}<0$ usually, then we can rewrite this as $|A_t|<\sqrt{6}
\msusy$, assuming $m_{\tilde t_L}=m_{\tilde t_R}=\msusy$. Thus we
might also expect that 
\beq |A_\kappa|<\sqrt{6}\msusy.
\label{ccb1}
\eeq
However, because of the presence of the $M_V$ mass term, Eq.~(\ref{ccb1})
generalizes to:
\beq
A_\kappa^2 < 6\left(M_V^2+\msusy^2\right)
\label{ccb2}
\eeq
which allows for much larger values of $A_\kappa$ and thus
$X_\kappa$. There are theoretical arguments both for and against
loosening the constraint of Eq.~(\ref{ccb1}) to that of
Eq.~(\ref{ccb2}). Since $M_V$ is {\it a priori}\/ independent of
$\msusy$, there is no good reason for a soft-breaking parameter like
$A_\kappa$ to scale with $M_V$. And in fact, for $M_V\sim 2\tev$,
Eq.~(\ref{ccb2}) would allow $A_\kappa$ as large as $5\tev$. Even if
such a large $A$-term were allowed phenomenologically, it would
generate a very large fine-tuning in the soft parameters of the Higgs
potential. On the other hand, it would be natural for $M_V$ to be
associated with the SUSY-breaking scale, just as the usual $\mu$-term
is presumed to be. So $A_\kappa$ scaling with $M_V$ may not be as
unreasonable as it first appears. In either case, we will remain
agnostic and show the limits under the constraints of 
Eqs.~(\ref{ccb1}) and (\ref{ccb2}) both.

\begin{figure}
\centering
\epsfclipon
\epsfysize=3.25in
\hspace*{0in}
\epsffile{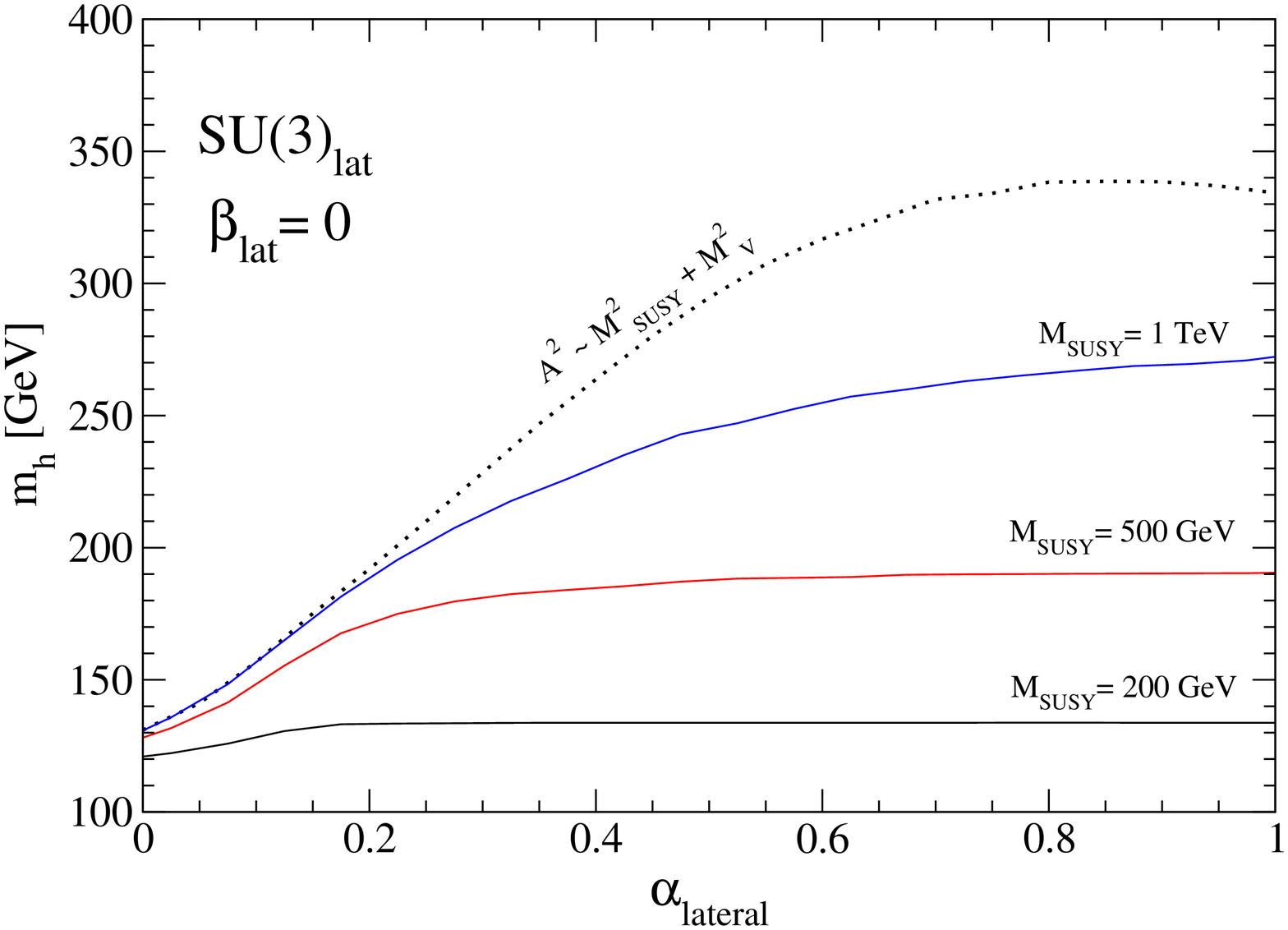}
\caption{Plot of maximum allowed Higgs mass, $m_h$, as a
  function of $\alpha_\lat$ for $SU(3)_\lat$ with $\beta=0$. 
The three lower lines assume $A^2\sim
  \msusy^2$ and $\msusy=200$, 500, and $1000\gev$ respectively. The
  topmost dotted line represents the limit if $A^2\sim \msusy^2+M_V^2$ (and
  is roughly indepedent of $\msusy$ itself).}
\label{plotone}
\vspace{0.75cm}
\epsfysize=3.25in
\hspace*{0in}
\epsffile{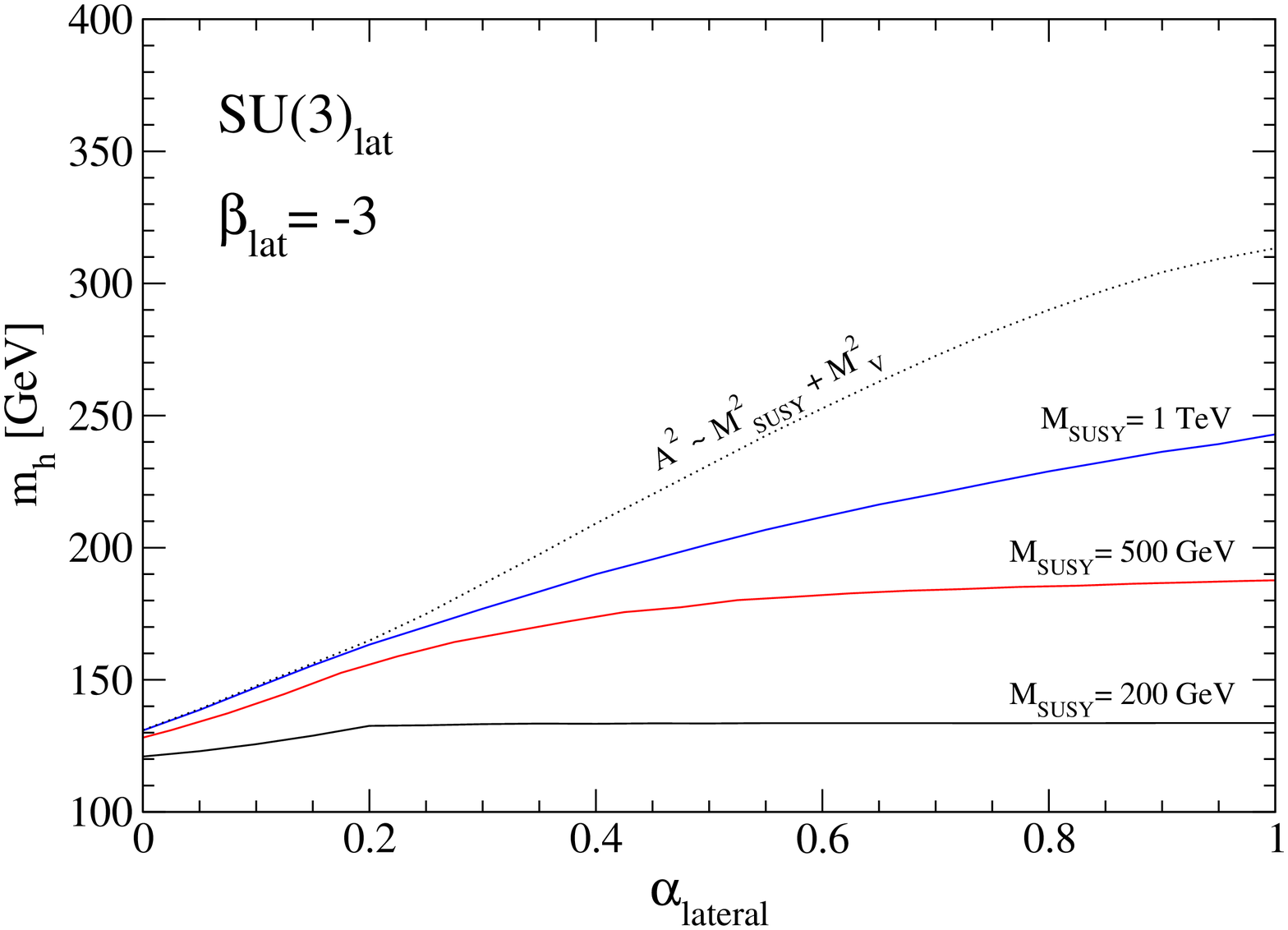}
\caption{Plot of maximum allowed Higgs mass, $m_h$, as a
  function of $\alpha_\lat$ for $SU(3)_\lat$ with $\beta=-3$. 
The lines follow those in Fig.~\ref{plotone}.}
\label{plottwo}
\end{figure}

We are now prepared to calculate the Higgs mass in our lateral $SU(3)$
model. In Fig.~\ref{plotone} we plot the maximum Higgs mass that can be
obtained as a function of $\alpha_\lat$ and for given values of
$\msusy$. The lower three lines in the figure 
assume that $A^2_\kappa\sim \msusy^2$ (that is, we impose
Eq.~(\ref{ccb1})). Each line represents a {\it maximum}\/ value of
$m_h$ consistent with a given $\msusy$. These points are maxima in
the following sense.
First, the stop mass
correction has been maximized by assuming $X_t=6$ for the stops. Second,
at each point along the curves, the vector-like mass, $M_V$, has been
allowed to vary within the range
\beq
500\gev \leq M_V \leq 2\tev
\label{mv}
\eeq
in order to maximize the size of the corrections.
The lower bound in Eq.~(\ref{mv})
has been chosen to satisfy the $T$-parameter
constraint, and the upper bound ensures that the vector-like matter
does not decouple completely.
The dependence of $m_h$ on 
$M_V$ is what one would expect from an examination of
Eq.~(\ref{hm2}). For $\alpha_\lat$
small, it is advantageous to have as large a logarithmic enhancement
as possible and so $M_V$ should be small. For $\alpha_\lat$ large, the
2-loop corrections push down $m_h$ if the log is large, and so it is
best to have a small log, which requires $M_V$ to be large.
One finds $M_V\approx 500\gev$ at $\alpha_\lat=0$ and $M_V\approx
2\tev$ when $\alpha_\lat\to1$. This effect is further enhanced by the
$T$-parameter constraint: larger $\alpha_\lat$ allow larger $\kappa$
which are consistent with $T$ only if $M_V$ is larger, and vice-versa.

Obviously there is a strong dependence on $\alpha_\lat$, which is to
be expected. As $\alpha_\lat$ grows, the running of $\kappa$ is
suppressed and so larger weak-scale values of $\kappa$ are consistent
with our perturbativity constraints. Of course, as $\alpha_\lat$ grows
the weak scale model itself becomes less and less
perturbative. However we feel that there should be no problem trusting
the calculation of the Higgs mass correction even at relatively large
values of $\alpha_\lat$. As a check, we have compared the size of the
one-loop corrections to those at two loops and find the two-loop
corrections to be uniformly small compared to the one-loop pieces,
usually only 5\% to 20\% of the latter. Although our calculation
appears to be trustworthy up to at least $\alpha_\lat\simeq1$, one need
only go up to $\alpha_\lat\simeq0.3$ in order to generate large
corrections to $m_h$.

We note a few interesting features in Fig.~\ref{plotone}.
For $\alpha_\lat=0$, one reproduces the upper bounds on the Higgs mass
in the MSSM: 121, 128 and $131\gev$ for $\msusy=200$, 500 and
$1000\gev$ respectively.
For $\alpha_\lat$ as small as $0.118$ (the value of $\alpha_s$ at
$m_Z$), one already finds the Higgs mass pushed up to 125, 150, and
$160\gev$ for those same values of $\msusy$. All of
these values are well above the current mass bounds. But it is as
$\alpha_\lat$ gets larger that the new matter really affects the
calculation. For $\alpha_\lat=0.3$, the bounds on $m_h$ rise to 134,
181 and $213\gev$. 

Why is there a large dependence on $\msusy$? One would naively expect
that this is due to the logarithmic dependence on $\msusy$ in the
radiative corrections, but that is not the case. Most of the
effect here is due to the left-right mixing of the new matter. But 
if we require $A_\kappa^2
\sim \msusy^2$ (Eq.~(\ref{ccb1})) the mixing cannot
be close to maximal unless $\msusy$ is large.
(Recall that extra $M_V^2$
suppression in the denominator of Eq.~(\ref{X}).) This is the dominant
source of the $\msusy$ dependence. 

There is one other line present in Fig.~\ref{plotone}. The topmost,
dotted line represents the limit on $m_h$ when $A^2$ is allowed to
scale with $\msusy^2+M_V^2$. Now the
mixing can become maximal and $X_\kappa$ can approach the magic value
of 6. In this case, there is almost no depedence on $\msusy$ and so we
show the bound as a single line. Notice that for $\alpha_\lat \lsim 0.2$,
there is no real difference between the smaller and larger values of
$A_\kappa$. But for $\alpha_\lat=0.3$ the Higgs mass bound has
been pushed up from 213 to $230\gev$ by the larger $A_\kappa$ values.

Hidden in these plots is the dependence of our result on the lateral
gauge $\beta$-function. If we take $\beta=0$ we are in fact maximizing
the Higgs mass. We can see this for one particular case in
Fig.~\ref{plottwo} where we plot the maximal values of $m_h$ as a
function of $\alpha_\lat$ just as in Fig.~\ref{plotone}, but now with
$\beta_\lat=-3$ (the minimal model case). Notice that these bounds are
lower than those for $\beta_\lat=0$, particularly at large
$\tan\beta$. However, the difference is not great, and we would claim
that the $\beta=-3$ model is also successful at lifting the Higgs mass
well above the experimental limits.

\section*{Conclusions}

The famous upper bound on the Higgs mass in the MSSM may turn out 
to be one of the major stumbling blocks of the model and one of
our best clues on how to extend it. While it is relatively simple to
extend the MSSM to push the upper bound on $m_h$ up a few ten's of
GeV, larger changes usually require radical alterations of the model
which do not allow automatic GUT-like unification and suppression of
dimension-6 proton decay operators. 
We have shown here that it {\it is}\/ possible to extend
the MSSM in a relatively minimal way and still obtain Higgs boson
masses as heavy as 200 to $300\gev$. And we have done so while
preserving gauge coupling unification at scales around $10^{16}\gev$.

The theoretical price to be paid depends on the enhancement desired. For
the largest masses, one needs either large $\msusy$ or
large $L'$--$\bar S$ mixing through the $A_\kappa$ term. Either choice
reintroduces the ``little hierarchy'' problem since the low-energy
Higgs potential is sensitive to these mass scales. There is also a
``$\mu$-problem'' associated with setting $M_V$ near the weak scale,
but this is presumably solved in the same way that the usual
$\mu$-problem is solved. But most importantly, these models
are phenomenologically consistent and do not require any radical
changes in the MSSM or the ``grand desert'' scenario. 

For smaller $m_h$
enhancements of $50\gev$ or so, there seems to be little to
no price in terms of additional fine-tunings. In fact, the precision
electroweak data strongly prefers Higgs masses which are less than 200 to
$250\gev$~\cite{pdg}.
Thus we would argue that
this model would be a strong contender were the Higgs mass bound to
continue increasing or if the Higgs were to be actually found at
masses inconsistent with the MSSM but not heavier than about
$300\gev$.

\section*{Acknowledgments}
We would like to thank J.~Lennon for valuable discussions. 
The work of KB is supported in part by the US Department of Energy 
under grants DE-FG02-04ER46140 and DE-FG02-04ER41306, 
and by an award from the Research Corporation.
The work of IG and CK is supported in part by the National Science
Foundation under grant PHY00-98791.

\end{document}